\documentstyle[12pt]{article}

cm \oddsidemargin 0.5 true cm \evensidemargin 0.5 true cm

\newcommand\re[1]{{(\ref{#1})}}
\newcommand{\de}{\delta}

\def\e{\epsilon}

\newcommand{\be}{\begin{equation}}
\newcommand{\ee}{\end{equation}}
\newcommand{\bea}{\begin{eqnarray}}
\newcommand{\eea}{\end{eqnarray}}

\newcommand{\ba}{\begin{array}{c}}
\newcommand{\ea}{\end{array}}

\newcommand{\ve}{\varepsilon}

\newcommand{\pa}{\partial}

\begin{document}
\begin{flushright}
\vspace{1mm} hep-th/0110056                                                                                                                            \\
FIAN/TD/14/01\\
October 2001\\
\end{flushright}
\vspace{10mm}

\begin{center}
{\bf Point particle in general background fields vs.\\
free gauge theories of traceless symmetric tensors} \\

\vspace{10mm} Arkady Yu. Segal ${}^{\dag\ddag}$\\

\vspace{5mm}{$\dag$ \it Spinoza Institute, Utrecht, The
Netherlands}

 \vspace{3mm}{$\ddag$ \it I.E.Tamm Department of Theoretical
Physics,
Lebedev Physical Institute,\\Leninsky prospect 53, 119991, Moscow, Russia\\
e-mail: segal@lpi.ru  }

\vspace{2mm}

\end{center}
\vspace{5mm}
\begin{abstract}
Point particle may interact to traceless symmetric tensors of
arbitrary rank. Free gauge theories of traceless symmetric
tensors are constructed, that provides a possibility for a new
type of interactions, when particles exchange by those gauge
fields. The gauge theories are parameterized by the particle's
mass $m$ and otherwise are unique for each rank $s$. For $m=0$,
they are local gauge models with actions of $2s$-th order in
derivatives, known in $d=4$ as "pure spin", or "conformal higher
spin" actions by Fradkin and Tseytlin. For $m\neq 0$, each
rank-$s$ model undergoes a unique nonlocal deformation which
entangles fields of all ranks, starting from $s$. There exists a
nonlocal transform which maps $m \neq 0$ theories onto $m=0$
ones, however, this map degenerates at some $m\neq 0$ fields
whose polarizations are determined by zeros of Bessel functions.
Conformal covariance properties of the $m=0$ models are analyzed,
the space of gauge fields is shown to admit an action of an
infinite-dimensional "conformal higher spin" Lie algebra which
leaves gauge transformations intact.
\end{abstract}

%
%
%

\newpage
\section{Introduction and background. \label{s0}}    

\vspace{5mm} The study of point particles living in background of
electromagnetic and gravitational fields represented by rank-$1$
and rank-$2$ tensor fields, is basic in General Relativity. On the
other hand, point particles may experience an influence of higher
rank symmetric tensors.

In 1980, De Wit and Freedman had addressed the generalization of
the point particle dynamics in gravitational and electromagnetic
background to the case when higher rank symmetric tensors are
switched on \cite{DeWitFreedman}. Specifically, they had
considered the action\footnote{On our notation: we use signature
$(-++...+)$, alternative as compared to DeWit and Freedman's paper
and re-introduce the mass parameter $m$ explicitly. The action
\re{hamBdW1} coincides with the De Wit-Freedman ($DW-F$) one
\cite{DeWitFreedman} after the identification
$\varphi^{DW-F}_{m_1...m_k} = - h_{m_1...m_k}$ and setting
$m^2=-1$ (the negative sign of $m^2$ just accounts the difference
in metric's signature,
$\sqrt{\dot{x}^2}{}|_{DW-F}=\sqrt{-m^2\dot{x}^2}|_{our})$.}
 \be \label {hamBdW1}
S\,[x(\tau)\,|\,h_{m(k)}] =- \int d\tau \{ \sqrt{-m^2\dot{x}^2}
\left(1+ \frac{e_k}{m^2} h_{m_1...m_k}(x)\,
\dot{x}^{m_1}...\dot{x}^{m_k}
(-\frac{m^2}{\dot{x}^2})^{\frac{k}{2}} \right)\}, \ee where
$x^m(\tau),\,m=0,1,...d-1$ represent the particle's world-line,
$h_{m_1...m_k}(x)$ are symmetric tensor fields and $e_k$ are
corresponding coupling constants, while $\dot{x}^2=g_{mn}(x)
\dot{x}^m \dot{x}^n$ with a general metric $g_{mn}(x)$. The
action is clearly invariant under world-line reparametrizations
and thereby governs some good point particle's space-time
evolution.

It was demonstrated also that the action possesses the {\it
first-order} invariance w.r.t following simultaneous
transformations of background fields $h_{m_1...m_k}(x)$ and
particle's world lines\footnote{Whenever indices denoted by the
same letter appear in our paper, their full symmetrization is
implied, the symbols like $m(s)$ stand for $m_1...m_s$.}:
\begin{equation}\label{DWFtrans}
\begin{array}{c}
S\,[x+\delta x \,|\, h_{m(k)}+ \delta h_{m(k)}]
=S\,[x \,|\,h_{m(k)}] + o(e_k^2)\\ \\
\delta h_{m(k)}(x)=\epsilon_{m(k-1);\,m} \\ \\
\delta x^m (\tau)=(k-1)\, e_k \,{\epsilon^{m}}_{n_1
...n_{k-2}}\dot{x}^{n_1}...\dot{x}^{n_{k-2}}\,({-\frac{\dot{x}^2}{m^2}})^{1-k/2},
\end{array}
\end{equation}
where $\epsilon_{m(k-1)}(x)$\footnote{The parameters
$\epsilon_{m(k-1)}$ correspond to gauge parameters $\xi_{m(k-1)}$
of \cite{DeWitFreedman} as $\epsilon_{m(k-1)}=k\xi_{m(k-1)},
k=1,2,...$} are arbitrary functions of $x^m$, $";"$ denotes the
covariant derivative compatible with the metric $g_{mn}$.

For $k=1,2$, one gets the standard gauge transformations for the
fluctuations of Maxwell and gravitational fields
\begin{equation}\label{lspinDWF}
\delta h_{m}(x)= \epsilon_{,\,m}\;\;,\;\; \delta
h_{m(2)}(x)=\epsilon_{m;\,m},
\end{equation}
so, if the higher ($k>2$) fields are set zero, the action
\re{hamBdW1} may be viewed as that describing the first-order
interaction of the particle to general fluctuations of
gravitational and Maxwell fields.

Clearly, the $k=1,2$ transformations  for background fields
\re{lspinDWF} present linearization of full $U(1)$ and general
coordinate ones. A natural question is then what is a nonlinear
generalization of linearized higher-rank transformations
\re{DWFtrans}. This question has recently been answered in the
author's paper \cite{Segal:2000ke}. It appears that, if one
passes to Hamiltonian formalism, the action \re{hamBdW1} turns
out to present the first order approximation to the general
Hamiltonian action of the point particle (without higher
derivatives), while the symmetries \re{DWFtrans} are nothing but
canonical transformations of the particle's phase space, with
higher order terms thrown away, this automatically insures the
invariance of the action \re{hamBdW1}\footnote{Also, the
Hamiltonian treatment automatically implies equality of all the
coupling constants, $e_k=e$. }. The nonlinear invariance
identified in \cite{Segal:2000ke} is a semidirect product of all
canonical transformations to an abelian ideal of "hyperWeyl"
transformations. The origin of these "hyperWeyl" transformations
is due to the fact being exposed already in the first order action
\re{hamBdW1}: the trace-part of every rank-$k$ tensor provides
exactly the same contribution to the particle action as a
rank-($k-2$) tensor. So, if one considers interaction of the
particle to \textit{all} symmetric tensors altogether, only
traceless parts of rank-$k$ tensors are actually involved. Below
we will provide, following \cite{Segal:2000ke}, a brief
derivation of all these matters in the Hamiltonian formalism,
which, in our opinion, most simply incorporates all essential
properties of the model and, besides, includes naturally the
massless case $m=0$.

The second natural question is whether there exist some dynamical
equations, consistent with linearized gauge invariance, for
higher rank background fields (that is something analogous to
Einstein and Maxwell equations for low rank tensors). If such
equations do exist \textit{at least at the first-order level}
then it becomes possible to consider first order processes like
"the particle "A" emits a higher rank field "F" which propagates
through the space-time according to its linearized field
equations and then hits the particle "B"", and thereby point
particle would interact to each other by means of higher rank
symmetric tensors.

\textbf{In this paper,} we address this second question, in
arbitrary space-time dimension $d$. Our starting point is the
first-order gauge transformations (with the \textit{flat} metric
background) for the infinite system of symmetric traceless
tensors derived in \cite{Segal:2000ke}. Our target is a gauge
invariant and Poincar\'e invariant quadratic action for these
fields. We show that, for $m=0$, for each rank $s$, there exists a
local action of $2s$-th order in derivatives, which is
\textit{unique} (modulo fields redefinitions of a trivial type)
and scales homogeneously under dilations. We call it spin-$s$
\textit{traceless higher spin theory}. In $d=4$, this model
coincides with "pure spin-$s$ model", described by Fradkin and
Tseytlin and conjectured to be conformally invariant \cite{FT}.
These higher derivative models should not be confused with second
order Fronsdal higher spin theories \cite{fronsdal1} described in
terms of \textit{double-traceless} tensors.

For $m \neq 0$, each traceless higher spin-$s$ theory undergoes a
deformation to a \textit{unique} and \textit{nonlocal} one (with
the nonlocality governed by the $\frac{m^2}{\Box}$ operator),
which mixes fields of all ranks, starting from $s$, and reduces to
corresponding local traceless higher spin theory at the point
$m=0$. We have found a nonlocal transform that maps $m \neq 0$
models to $m=0$ ones, however it degenerates at some $m \neq 0$
fields which may be of importance. Therefore, it may be worth
studying these theories in the original basis, where they are
nonlocal, and name them differently from $m=0$ models. We will
call these deformed models \textit{"deformed traceless higher
spin theories"}. According to the above reasoning, all these
theories are of interest, as they can govern unique first order
interactions of either massless point particles via traceless
higher spin fields or massive point particles via deformed
traceless higher spin fields.

The paper is organized as follows. In Section \ref{s1}, we briefly
re-derive, following \cite{Segal:2000ke}, linearized gauge
transformations for the infinite collection of symmetric tensors,
governing first-order dynamics of the point particle. In Section
\ref{s2}, we derive the quadratic gauge invariant actions both for
$m=0$ and $m \neq 0$ case and analyze the nonlocal map from
$m\neq 0$ to $m=0$. In Section \ref{s3}, we analyze some
conformal covariance properties of $m=0$ models, specifically, we
show that the space of gauge fields may be assigned with an
action of an infinite-dimensional "conformal higher spin" Lie
algebra (which contains the conformal algebra as its maximal
finite-dimensional subalgebra) in such a way that gauge
transformations remain intact. In Conclusion, we discuss the
results and outline a possibility of studying the nonlinear
action.

\section{Point particle and gauge transformations for symmetric traceless
tensors.}\label{s1}

The Hamiltonian action of a point particle in general background
fields reads \be \label {ham} S_H [x(\tau), p(\tau),
\lambda(\tau)]=\int d\tau \{p_m \dot{x}^m -\lambda H (p, q)\},
\ee where $x^m(\tau), m=0,1...d-1,$ are the coordinates of the
particle's world line, $p_m(\tau)$ are the momenta and $\lambda$
is a Lagrange multiplier to the unique first class constraint $H
(x^m, p_m) \approx 0 $ which we shall call Hamiltonian. The
Hamiltonian is supposed to be a power series in momenta, \be
\label{hclass} H = \sum\limits_{k=0}^\infty H^{m_1...m_k} (x)
p_{m_1}...p_{m_k} = \sum\limits_{k=0}^\infty H_k \ee where $H_k$
denotes the homogeneous polynomial of $k$-th degree in momenta.
When $H_k=0$ for $k>2$, the model describes a particle in general
gravitational + Maxwell background, while otherwise the particle
experiences the influence of  higher rank symmetric tensors
$H^{m_1...m_k}(x)$. In \cite{Segal:2000ke}, the gauge
transformations for $H^{m_1...m_k}(x)$ were derived by
postulating that Hamiltonians $H$ and $H'$ are gauge equivalent if
they describe equivalent particle's dynamics\footnote{It should be
noted that the notion of \textit{physical} equivalence is not
straightforward, however, the "generalized equivalence principle"
which underlies our derivation \cite{Segal:2000ke}, is relevant
at least because it automatically provides one with closed gauge
transformations as forming the covariance algebra of some physical
system (a particle).}. Specifically, if one makes an infinitesimal
canonical transformation $x'(x,p)=x+\delta x$, $p'(x,p)=p+ \delta
p$,
\begin{equation}
\label{invar} \delta{x^m}=\{x^m, \epsilon\}, \qquad \delta{p_m}=
\{p_m, \epsilon\},
\end{equation}
with generating function $\epsilon(x,p)$ ($\{ , \}$ stands for
the canonical Poisson bracket, $\{x^m, p_n\}
=\delta^m_n\,,\,\{x^m, x^n\}=\{p_m, p_n\}=0$), the dynamics in
$x',p'$ variables is determined by the canonically transformed
Hamiltonian \be\label{canon}\begin{array}{c}
H'(x,p)=H(x,p)+\delta H(x,p) \\ \\
\delta H(x,p) = \{\epsilon, H(x,p)\},\end{array} \ee which is, by
definition, equivalent to $H$.
 Provided $\epsilon(x,p)$ is also a power series in
momenta, \be \label{eclass} \epsilon = \sum\limits_{k=0}^\infty
\epsilon^{m_1...m_k} (x) p_{m_1}...p_{m_k} =
\sum\limits_{k=0}^\infty \epsilon_k, \ee these transformations
provide the action of the canonical transformations algebra on
the infinite collection of symmetric tensor fields $H^{m_1...m_k}
(x)$ comprising the power series $H(x,p)$. Also, the Hamiltonians
differing by a factor which never comes to zero,
\begin{equation}\label{hyperweyl}
 H'(x,p)= A(x,p)\, H(x,p)\,,\,A(x,p)\neq 0
\end{equation}
are by definition equivalent (they do determine the same dynamics
of the particle, as its dynamics is localized on the constraint
surface $H\approx0$). Representing $A(x,p)$ as $A=e^{a(x,p)}$, one
may write down the infinitesimal form of \re{hyperweyl}
\begin{equation}\label{hyperweyl1}
\delta H(x,p)= a(x,p) H(x,p),
\end{equation}
where $a(x,p)$ is also a power series in momenta: \be
\label{aclass} a= \sum\limits_{k=0}^\infty a^{m_1...m_k} (x)
p_{m_1}...p_{m_k} = \sum\limits_{k=0}^\infty a_k. \ee  As a
result, one gets the action of some huge gauge algebra,
\begin{equation}\label{total}
\delta_{(a,\epsilon)} H(x,p)= a(x,p) H(x,p)+\{ \epsilon,H(x,p)\}
\end{equation}
 on the
infinite collection of symmetric tensor fields $H^{m_1...m_k}
(x)$. This algebra clearly contains $U(1)$ "phase" transformations
and $x$-diffeomorphisms, generated by $\epsilon
=\varepsilon(x)+\xi^m (x)p_m$ ($\varepsilon$ generates $U(1)$ and
$\xi^m$ the diffeomorphisms), and Weyl dilations, generated by
$p$-independent $a(x,p)=a_0$, and contains much more.

Expand $H(x,p)$ around the natural vacuum
\begin{equation}\label{Hvac}
H=H_v+h(x,p) \equiv \frac{1}{2}(\eta^{mn}p_{m} p_{n} + m^2) +
h(x,p),
\end{equation}
with Minkowski background for metric $g_{mn}=\eta_{mn}$. It is
worth noting that after passing to the Lagrangian formulation and
throwing away the higher (except zero and first) orders in $h$ in
the particle's Lagrangian, one arrives exactly at the sum of the
actions \re{hamBdW1} \cite{Segal:2000ke}.

Now rewrite the gauge transformations \re{total} in terms of
$h(x,p)$ and make the linearization, i.e. extract the lowest
order in $h(x,p)$. On obtains
\begin{equation}\label{lintotal}
\delta h(x,p)=a(x,p) H_v+\{ \epsilon,H_v\} \equiv \frac{1}{2}
a(x,p)(p^2+m^2) + p_m \eta^{mn} \partial_n \epsilon (x,p),
\end{equation}
where $\partial_{m}$ is the derivative w.r.t. $x^m$. In the
component form, the gauge transformations read
\begin{equation}\label{lintotal1}
  \delta h^{m(s)} =\frac{1}{2}( \eta^{m(2)} a^{m(s-2)} +m^2 a^{m(s)}) + \partial^m
  \epsilon^{m(s-1)}.
\end{equation}
Our program is to look for  quadratic theories, invariant w.r.t
these gauge transformations. Before dwelling on details let us
note that these theories will describe dynamics of infinite
collection of \textit{traceless} tensors, as the trace parts of
$h^{m(s)}$ are gauged away by purely algebraic
$a(x,p)$-transformations. It is worth extracting the invariants of
$a$-transformations and looking how these invariants transform
under $\epsilon$-gauge symmetries.

The simplest case is $m^2=0$ one. Representing
$h^{m(s)},\epsilon^{m(s-1)}$ as a sum of their traceless and trace
parts,
\begin{equation}\label{exhwt1}
 \begin{array}{c}
   h^{m(s)}=\varphi^{m(s)} + \eta^{m(2)}\,\chi^{m(s-2)}\;;\;{\varphi_l}^{lm(s-2)} =0, \\
   \epsilon^{m(s-1)}=\varepsilon^{m(s)} + \eta^{m(2)}\, \zeta^{m(s-3)}\;;\;{\varepsilon_l}^{lm(s-3)}
   =0,
 \end{array}
\end{equation}
one observes that all the traces $\chi^{m(s-2)}$ are gauged away
by $a$-transformations, while for the traceless parts one derives
the gauge transformations
\begin{equation}\label{exhw2}
\delta \varphi^{m(s)}={\mbox{Traceless part of}}\;\pa^m
\varepsilon^{m(s-1)}.
\end{equation}
The invariant action will be shown below to present
scale-covariant theory of $2s$-th order in derivatives, formulated
in terms of the tracelss tensor of rank $s$.

In the $m \neq 0$ case, the situation is more complicated. To
describe it, we repeat the derivation from \cite{Segal:2000ke} in
the Appendix. The result is that, in the $m^2 \neq 0$ case,
$a$-invariants are traceless tensors $\varphi^{m(s)}$ \re{5} built
out of $h^{m(k)}$, which possess gauge transformations \be
\label{gtra1}  \de \varphi^{\,m(s)} =( {\mbox{Traceless part
of}}\; \pa^{m} \ve^{m(s-1)}) -m^2\, \frac{s+1}{2s+d}\, \pa_n
{\ve^{nm(s)}}.\ee These gauge transformations entangle components
of all ranks, and in this case the invariant actions will be
shown to be nonlocal and involve fields of all ranks from $s$ to
$\infty$.

\section{Invariant actions.} \label{s2}
Now let us look for a Poincar\'e- and gauge-invariant action. The
most general Poincar\'e-invariant quadratic action is
\begin{equation}\label{Saction}
  A_P[h]=\sum \limits_{k=0,k'=0}^{\infty} \int d^d x
  h^{m_1...m_k}(x) P_{\{m_1...m_k | n_1...n_{k'}\}} (\partial_l)h^{n_1...n_{k'}} (x),
\end{equation}
where $P_{ \{ m_1...m_k | n_1...n_k \} } (\partial_l)$ are some
(pseudo)differential operators constructed from the partial
derivative $\partial_m$ and the Minkowski metric, they are also
allowed to contain any function of $\Box$. The operator $P_{ \{
m(k)| n(k) \} } (\partial_l)$ is supposed to be symmetric,
therefore
\begin{equation}\label{Psym}
P_{ \{ m(k)| n(k') \} } (-\partial_l)=P_{ \{ n(k')| m(k) \} }
(\partial_l).\end{equation} The gauge invariance of the quadratic
action is equivalent to the gauge invariance of the equations of
motion
\begin{equation}\label{Siden}
\sum \limits_{k'=0}^{\infty} P_{\{m_1...m_k | n_1...n_{k'}\}}
(\partial_l)  h^{n_1...n_{k'}}(x)= 0 ; \forall k.
\end{equation}
To find out the solution for $P$'s one has to substitute the
transformations \re{lintotal1} into the last equation and require
all the identities associated with a given parameter ($a^{m(k)}$
or $\epsilon^{m(k)}$) to hold. It is easy to see $P$'s have to be
$\pa_m$-transversal and satisfy certain tracelessness constraints:
\begin{equation}\label{constr}
\begin{array}{c}
   P_{\{l_1,l_2...l_k | n_1...n_{k'-1}r\}}\,\pa^r =0; \\ \\
 \eta^{ab} P_{\{l_1...l_k | n_1...n_{k'-2}a\,b\}} +m^2 P_{\{l_1...l_k |
n_1...n_{k'-2}\}}=0;\,\forall k,k',
\end{array}
\end{equation}
and analogously for $n\leftrightarrow l$.

It turns out this infinite system of identities may be studied in
a simple generating framework. Introduce the power series of two
variables $q^m, q'^m$
\begin{equation}\label{generP}
P(q,q')=\sum \limits_{k=0,k'=0}^{\infty} \frac{1}{k!k'!}
q^{n_1}...q^{n_{k'}} q'{}^{m_1}...q'{}^{m_{k}} P_{\{m_1...m_k |
n_1...n_{k'}\}} (\partial_l).
\end{equation}
Then the infinite system of identities \re{Siden} is equivalent
to two equations on $P(q,q')$:
\begin{equation}\label{baseq1}
\nabla\, P(q,q') =0\,;\,\nabla \equiv \eta^{mn}
\frac{\partial}{\partial q^m} \frac{\partial}{\partial x^n}
\end{equation}
and
\begin{equation}\label{baseq2}
( \diamond +m^2)\,P(q,q' ) =0 \,;\, \diamond \equiv \eta^{mn}
\frac{\partial}{\partial q^m} \frac{\partial}{\partial q^n}.
\end{equation}
The equation \re{Psym} turns into
\begin{equation}\label{Psym1}
P(q,q')(\partial)=P(q',q)(-\partial),
\end{equation}
and therefore one gets also
\begin{equation}\label{baseq3}
\nabla'\, P(q,q') =0\,;\,\nabla' \equiv \eta^{mn}
\frac{\partial}{\partial q'^m} \frac{\partial}{\partial x^n}
\end{equation}
and
\begin{equation}\label{baseq4}
( \diamond' +m^2 )\,P(q,q' ) =0 \,;\, \diamond' \equiv \eta^{mn}
\frac{\partial}{\partial q'^m} \frac{\partial}{\partial q'^n}.
\end{equation}
The equations \re{baseq1}-\re{baseq4} constitute the full set of
conditions for gauge invariance of the action \re{Saction}.

Before starting to solve the equations \re{baseq1}-\re{baseq4},
it is worth noting the important property of the formalism. If one
"dresses" $P(q,q')$ like
\begin{equation}\label{dress}
  P(q,q')=U(q)U(q')P_U(q,q') \,\Leftrightarrow\,
  P_U(q,q')=U^{-1}(q)\,U(q')^{-1}\,P(q,q'),
\end{equation}
with some $U(q)\neq 0$, then new operator $P_U(q,q')$ already
does not satisfy the same equations for $P(q,q')$. However, if
this change is accompanied by the following "dressing" of gauge
fields
\begin{equation}\label{dress1}
h(x,p)=U(\frac{\pa}{\pa p}) h_U(x,p) \,\Leftrightarrow
h_U(x,p)=U^{-1}(\frac{\pa}{\pa p })h(x,p)
\end{equation}
then it results in the same theory, i.e.
\begin{equation}\label{dress2}
 A_{P_U} [ h_U ] = A_P[h].
\end{equation}
This property of "dressing" will play an important role below.

 Now let us find the solution to these equations. Note that
the partial derivative w.r.t $x^m$ enters all these equations
just as some constant vector, to emphasize this we will denote it
below as some constant vector
$$
\frac{\partial}{\partial x^n}\equiv d_n.
$$
Below, we will also denote $d_m d^m \equiv \Box$ and use the
self-evident notation like $(AB) \equiv A_k B^k, A^2 \equiv A_k
A^k$.

First, we study the equations \re{baseq1} and \re{baseq3}. Their
(Poincar\'e-covariant) general solution is easily obtained as
they are just first order differential equations w.r.t. $q$, $q'$:
$P(q,q')$ should depend on the $d^n$-transversal combinations:
\begin{equation}\label{transv}
 \begin{array}{c}
  P=P(q_{\bot},q'_{\bot})  \\ \\
q^m_{\bot} = q^m -\frac{(qd)d^m}{\Box};\\ \\{q'}^m_{\bot} = {q'}^m
-\frac{(q'd)d^m}{\Box}. \end{array}
\end{equation}
As the solution has to be Poincar\'e invariant, $P$ should be a
function of three Poincar\'e invariants built out of
$q^m_{\bot},q'^m_{\bot}$:
\begin{equation}\label{threeinvs}
\begin{array}{c}
P=Q\,(\sigma,\sigma',\tau),\\ \\ \sigma =\Box \,q^2_{\bot} =\Box
\,q^2 -(qd)^2\;;\; \sigma' =\Box \,q'^2_{\bot} =\Box \,q'^2
-(q'd)^2
\\ \\ \tau = \Box \,(q_{\bot} q'_{\bot}) = \Box \,(qq') -(qd)(q'd).
\end{array}
\end{equation}
Now we have to determine the function of three variables $Q$. Note
that $\sigma, \sigma', \tau$ should enter the solution only
analytically, otherwise the solution could not be interpreted as
power series in $q,q'$. By virtue of \re{Psym1} $Q$ satisfies
\begin{equation}\label{Psym2}
Q(\sigma,\sigma',\tau)=Q(\sigma',\sigma,\tau).
\end{equation}
Implement the equation \re{baseq2}. We get (commas in subscripts
denote derivatives w.r.t. $q^m$)
\begin{equation}\label{baseq22}
\begin{array}{c}
0= ( \diamond +m^2 )\,Q = m^2 Q+ \\ \\ +\frac{\partial^2
Q}{\partial \sigma^2}\, \sigma_{,m} {\sigma,}^{m}
+\frac{\partial^2 Q}{\partial \sigma'^2}\,\sigma'_{,m}
{\sigma',}^{m} +\frac{\partial^2
Q}{\partial \tau^2} \,\tau_{,m} {\tau,}^{m} + \\ \\
+2 \frac{\partial^2 Q}{\partial \sigma \partial \tau}
\,\sigma_{,m} {\tau,}^{m}+2 \frac{\partial^2 Q}{\partial \sigma'
\partial \tau} \,\sigma'_{,m} {\tau,}^{m}+2 \frac{\partial^2
Q}{\partial \sigma
\partial \sigma'} \, \sigma_{,m} {\sigma',}^{m}+\\ \\
+ \frac{\partial Q}{\partial \sigma}\, {\sigma,^m}_{m} +
\frac{\partial Q}{\partial \sigma'} \,{\sigma',^m}_{m}+
\frac{\partial Q}{\partial \tau}\, {\tau,^m}_{m}.
\end{array}
\end{equation}
After employing the identities
\begin{equation}\label{ids}
\begin{array}{c}
\sigma_{,m} {\sigma,}^{m}=4 \Box \,\sigma\,;\, \tau_{,m}
{\tau,}^{m} =\Box \,\sigma'\,;\, \sigma_{,m}
{\tau,}^{m}=2 \Box \,\tau\\ \\
{\sigma,^m}_{m}= 2 \Box \,(d-1)\,;\,
{\tau,^m}_{m}=0\,;{\sigma',_m}=0,
\end{array}
\end{equation}
the equation \re{baseq22} is rewritten as
\begin{equation}\label{baseq222}
0= m^2 Q + \Box\,( 4\sigma \frac{\partial^2 Q}{\partial \sigma^2}
+\sigma'\, \frac{\partial^2 Q}{\partial \tau^2} +4\tau\,
\frac{\partial^2 Q}{\partial \sigma \partial \tau}
+2(d-1)\frac{\partial Q}{\partial \sigma}).
\end{equation}
It is convenient to represent $Q$ as a power series in $\tau$,
\begin{equation}\label{Qseries}
  Q=\sum \limits_{l=0}^{\infty} Q_l (\sigma, \sigma')
\tau^l,
\end{equation}
then the equation \re{baseq222} reads
\begin{equation}\label{baseq223}
 ( 4\sigma \frac{\partial^2 }{\partial \sigma^2}
+2(2l+d-1)\frac{\partial }{\partial \sigma}+\mu^2) Q_l +
(l+1)(l+2) Q_{l+2}\,\sigma' =0,
\end{equation}
where the "nonlocality" parameter $\mu^2 \equiv \frac{m^2}{\Box}$
is introduced. The $\sigma'$ counterpart of this equation holds
either due to \re{baseq4}:
\begin{equation}\label{baseq224}
 ( 4\sigma' \frac{\partial^2 }{\partial \sigma'^2}
+2(2l+d-1)\frac{\partial }{\partial \sigma'}+\mu^2) Q_l +
(l+1)(l+2) Q_{l+2}\,\sigma =0.
\end{equation}
Making change of variables
\begin{equation}\label{chv}
\rho=\sqrt{\sigma}\,,\,\rho'=\sqrt{\sigma'}
\end{equation}
and introducing new function $R(\rho,\rho', \tau)$ by the rule
\begin{equation}\label{baseq225}
Q(\sigma,\sigma',\tau)=(\rho\rho')^{-\frac{d-3}{2}} R(\mu\rho,\mu
\rho',\frac{\tau}{\rho\rho'} )\equiv
(\rho\rho')^{-\frac{d-3}{2}}\sum \limits_{l=0}^{\infty}
R_l\,(\mu\rho,\mu\rho') (\frac{\tau}{\rho\rho'})^l,
\end{equation}
one rewrites the equations \re{baseq223} and \re{baseq224} in the
form
\begin{equation}\label{bess}
  \begin{array}{c}
( B_{\rho}-(l+\frac{d-3}{2})^2 ) R_l = -(l+1)(l+2) R_{l+2}\\ \\(
B_{\rho'}-(l+\frac{d-3}{2})^2 ) R_l = -(l+1)(l+2) R_{l+2},
  \end{array}
\end{equation}
where
\begin{equation}\label{bessop}
  B_{\rho} \equiv \rho^2 \frac{\partial^2}{\partial \rho^2} +\rho\,\frac{\partial}{\partial
  \rho} + \rho^2
\end{equation}
is the operator governing Bessel's equation (and $B_{\rho'}$ is
the same operator acting on $\rho'$). Represent $R_l(\rho,\rho')$
as a series
\begin{equation}\label{bess2}
R_l(\rho,\rho') =\sum \limits_{\nu,\nu' \in \Omega}
R_{\nu,\nu',l}\, C_{\nu}(\rho) C_{\nu'} (\rho'),
\end{equation}
where $\Omega$ is some set of points in the complex plane,
$R_{\nu,\nu',l}$ are constants, and $C_{\nu}(\rho)$ is a solution
of Bessel's equation with index $\nu$
\begin{equation}\label{bessop1}
(  B_{\rho} -\nu^2) C_{\nu} (\rho) \equiv (\rho^2
\frac{\partial^2}{\partial \rho^2} +\frac{\partial}{\partial
  \rho} + \rho^2 -\nu^2) C_{\nu}(\rho) =0.
\end{equation}
Then the equations \re{bess} take the form
\begin{equation}\label{bess3}
  \begin{array}{c}
( \nu^2-(l+\frac{d-3}{2})^2 ) R_{\nu,\nu',l} = -(l+1)(l+2) R_{\nu,\nu',l+2}\\
\\( \nu'^2-(l+\frac{d-3}{2})^2 ) R_{\nu,\nu',l} = -(l+1)(l+2) R_{\nu,\nu',l+2}.
  \end{array}
\end{equation}
It is seen that these equations are split into independent chains
of simple iterative equations w.r.t. $l$, where each chain
corresponds to particular values of Bessel indices $\nu,\nu'$. It
is readily seen that only solutions with $\nu^2 =\nu'^2$ are
nonzero, so one has to study just one family of iterative chains
\begin{equation}\label{bess31}
  \begin{array}{c}
r_{\nu,l} \equiv R_{\nu,\nu,l},\\
\\( \nu^2-(l+\frac{d-3}{2})^2 ) r_{\nu,l} = -(l+1)(l+2)
r_{\nu,l+2}.
  \end{array}
\end{equation}
Now recall that we are looking for solutions \re{baseq225} which
are  \textit{real} and \textit{analytic at the origin} in
$\sigma,\sigma'$ and $\tau$. Hence we have to choose only those
solutions for $R$ which are real and such that
$\rho^{-\frac{d-3}{2}} C_{\nu} (\rho)$ possesses analytic
decomposition at the origin. This particularly implies (see
\re{bess31}) $\nu^2$ is real. Then, $\nu^2$ should be nonnegative
as otherwise the iteration from $r_l$ to $r_{l+2}$ governed by
the equations \re{bess31}, never stops, which will result (see
(\ref{baseq225},\ref{bess2})) in arbitrarily large negative
powers of $\rho\rho'$.

 Denote
\begin{equation}\label{ga}
  \gamma \equiv \frac{d-3}{2}.
\end{equation}
Below we consider the case $d \geq 3$ and hence $\gamma \geq 0$
and after discussing this general case we return to $d=2$, while
$d=1$ case is obviously a trivial one (in $d=1$,
$\rho=\rho'=\tau\equiv0$). As $\gamma \geq 0$ while
$\rho^{-\gamma} C_{\nu} (\rho)$ should be analytic, the solution
$C_{\nu}{(\rho)}$ should be regular at the origin. For
nonnegative $\nu^2$, among the solutions to Bessel's equation,
only the Bessel's functions of the first kind and of nonnegative
index $\nu$ are known to possess this property, therefore one has
to make the choice
\begin{equation}\label{bess4}
\nu \geq 0\;;\; C_{\nu}(\rho)=J_{\nu}(\rho)=\sum
\limits_{k=0}^{\infty} \frac{(-1)^k}{k!\,\Gamma(\nu+k+1)}\;
(\frac{\rho}{2})^{\nu+2k} ,
\end{equation}
where $J_{\nu}(\rho)$ is the Bessel's function of the
\textit{first} kind, these solutions to Bessel's equations are
known to possess regular behaviour at $\rho=0$:
\begin{equation}\label{bessas}
J_{\nu} (\rho) \approx_{\rho\rightarrow 0}
\frac{1}{\Gamma(\nu+1)} (\frac{\rho}{2})^{\nu} ,
\end{equation}
Then, analyzing the behaviour of solutions \re{baseq225} at the
origin and applying this analyticity requirement along with
\re{bess2} and \re{bessas}, one gets that $Q$ is analytic in
$\sigma,\sigma'$ \textit{iff} $\nu -\gamma-l$ is \textit{even
positive integer or zero}, therefore
\begin{equation}\label{bess5}
  \nu-l-\gamma =2t \geq 0,\;t=0,1,2,...
\end{equation}
which implies
\begin{equation}\label{bess6}
  \nu \geq \gamma\;\;,\;\;0 \leq l \leq \nu - \gamma.
\end{equation}
Now one has to study the chain \re{bess31} and to find the
solutions consistent with \re{bess5},\re{bess6}. Their existence
is not \textit{a priori} guaranteed, as in accordance with
\re{bess6} solutions should contain finite number  of terms in
$l$. Nevertheless, the solutions do exist for each integer and
half-integer $\nu$. Denote
\begin{equation}\label{spin}
        s= \nu-\frac{d-3}{2} =0,1,2,3,...=2n+\varsigma \;;\;
   n=0,1,2,3,...\,\;\,,\,\varsigma=0,1
\end{equation}
so $\varsigma$ accounts whether $s$ is even or odd.

Then the only solutions satisfying \re{bess31} \textit{and}
\re{bess5} are
\begin{equation}\label{sol}
  \begin{array}{c}
    r_{2k+\varsigma, \gamma+2n+\varsigma}=0, \;\; k < 0 \;or\; k > n ; \\ \\
    r_{2k+\varsigma,\gamma+2n+\varsigma}=\frac{(-1)^k}{(2k+\varsigma)!}
    \frac{(2n)!!
    (2n+2\gamma+2\varsigma+2k-2)!!}{(2n-2k)!!(2n+2\gamma+2\varsigma-2)!!}\;
    r_{\varsigma,\gamma+2n+\varsigma}\equiv\\ \\\equiv
    c_{2k+\varsigma,\gamma+2n+\varsigma}\;r_{\varsigma,\gamma+2n+\varsigma};\\
    \\
    k=0,1,2,...,n-1,n,
     \end{array}
\end{equation}
where
\begin{equation}\label{2fact}
  a!!\equiv a(a-2)(a-4)...2 \;or\; 1.
\end{equation}
$r_{\varsigma,\gamma+2n+\varsigma}$ is an arbitrary "constant",
i.e. arbitrary function of $\Box$, which is convenient to redefine
as follows
\begin{equation}\label{rnorm}
r_{\varsigma,\gamma+2n+\varsigma}=(\frac{\mu}{2})^{-2(\gamma+2n+\varsigma)}
\;\tilde{r}_{n,\varsigma}
\end{equation}
The solutions \re{sol} exist because the coefficients of recurrent
relations \re{bess31} become zero at the three lines in the
$(l,\nu)$ plane, $L_{right}:$ $\nu=l+\gamma$ and $L_{1}, L_{2}:$
$l=-1,-2$, then the chain of iterations along the line
$L_{iterate}$: $\nu=s+\gamma$ terminates at its right edge as it
crosses $L_{right}$ and at its left edge as it crosses one of
$L_{1}, L_{2}$. It may be easily verified that, if $\nu \geq 0$,
these are the only solutions that contain finite number of points
along $L_{iterate}$\footnote{All other solutions contain infinite
number of terms in $l$, either infinite both to the left and to
the right or starting from some finite point in the $(l,\tau)$
plane and then going to infinity. All these solutions are
inappropriate for our considerations in the paper.}. And they are
exactly within the domain prescribed by the condition \re{bess6}.

On the other hand, the equations \re{bess31} are invariant w.r.t.
change $\nu \rightarrow -\nu$. It means that there is one more
branch of solutions which contains finite number of points along
$L_{iterate}$, it is obtained from (\ref{bess2},\ref{bess4}) by
substitution $J_{\nu} \mapsto J_{-\nu}$. These solutions are
already non-regular at the origin, so they are not connected with
our main task, however they could be of some interest as they are
still polynomial in $\tau$ and so may correspond to some "not so
wild" models. For integer $\nu$, the map $J_{\nu} \mapsto
J_{-\nu}$ does not produce new solutions as $J_{n}=(-)^n J_{-n}$,
but if one has already given up regularity at the origin, one can
choose the second solution to the Bessel's equation, the Hankel
function $C_{n}=Y_n$.

Thus, we have the general solution for the equations
\re{baseq1}-\re{baseq4}, which is parameterized by $\mu$,
$n=0,1,2,3,...$ and $\varsigma=0,1$:
\begin{equation}\label{baseq10}
P_{\{\mu,\, n,\varsigma\}}(q,q')= (\rho\rho')^{-\gamma}
J_{2n+\gamma+\varsigma} (\mu\rho) J_{2n+\gamma +\varsigma}
(\mu\rho') \sum \limits_{k=0}^{n}
(\frac{\tau}{\rho\rho'})^{2k+\varsigma}r_{2k+\varsigma,
2n+\gamma+\varsigma}
\end{equation}
Note that the solution contains only even powers of $\rho,\rho'$
as it should.

Let us analyze the theories we have obtained and clarify the
meaning of the parameters $n$ and $\varsigma$. To this purpose it
is worth taking the limit $m^2 \rightarrow 0$, or, what is the
same, $\mu \rightarrow 0$. In this case, only first term of the
Bessel's series survives,  so one gets, according to
\re{bessas}-\re{baseq10},
\begin{equation}\label{baseq11}
 P_{ \{0,\,n,\varsigma \} }(q,q')= \frac{{\tilde r}_{n,\varsigma}}{(\Gamma(2n+\varsigma
+\gamma+1))^2}\; \sum \limits_{k=0}^{n} {\tau}^{2k+\varsigma}\;
(\rho\rho')^{2(n-k)} c_{2k+\varsigma,\gamma+2n+\varsigma}.
\end{equation}
As it is clear from the very definition \re{generP}, a term
$I_{a,b,c}\sim \tau^a \rho^{2b} \rho'^{2c}$ in the solution leads
to the corresponding operator in the action \re{Saction}, which
contains $2a+2b+2c$ $x$-derivatives, and involve tensor fields of
ranks $a+2b$ and $a+2c$. Applying this account to the last
equation we see that the corresponding theory contains terms
$I_{2k+\varsigma,n-k,n-k}$ for $k=0,...,n$ and thereby describes
theory of symmetric tensor fields of rank-($2n+\varsigma$)
\textit{only}, with operators of \textit{only}
($2(2n+\varsigma)$) order in $x$-derivatives (for $n=\varsigma=0$,
the solution is just a constant, i.e. arbitrary function of
$\Box$). Therefore, these theories are covariant under scale
transformations $x'^m=e^\lambda x^m$. We call these models
\textit{traceless higher spin theories of spin $s$}. In $d=4$,
they were described by Fradkin and Tseytlin \cite{FT} ("pure
spin" models), and were conjectured to be invariant w.r.t. full
conformal algebra $so(4,2)$. They possess supersymmetric
extensions, studied by Fradkin and Linetsky \cite{FL1, FL2}.

In the next section, we will analyze some properties of traceless
higher spin models w.r.t. full conformal algebra in $d$
dimensions. We will show that if one considers all these models
altogether, the space of fields $h(x,p)$ may be assigned with an
action of an infinite-dimensional "conformal higher spin" Lie
algebra (which contains the conformal algebra as its maximal
finite-dimensional subalgebra) in such a way that gauge
transformations remain intact.

As $\mu\neq 0$, the solution presents a deformation of traceless
higher spin theories. These deformed models are \textit{nonlocal}
as the solution contains all the powers of
$\mu^2=\frac{m^2}{\Box}$ inside Bessel's functions. Moreover, as
the product of Bessel's functions of $\rho,\rho'$ is not the
product $\rho\rho'$, the corresponding theory seems to mix
tensors of all ranks from $2n+\varsigma$ to $\infty$. However, it
appears all these peculiar properties may be cancelled after a
change of variables. Indeed, the structure of solution
\re{baseq10} exhibits "dressing" \re{dress}:
\begin{equation}\label{dress4}
  P_{\{\mu,\,n,\varsigma\}}(\rho,\rho') =U_{2n+\varsigma+\gamma}(\mu \rho)\, U_{2n+\varsigma+\gamma}
  (\mu\rho')\, P_{\{0,\,n,\varsigma\}}(\rho,\rho'),
\end{equation}
where
\begin{equation}\label{dress5}
  U_{\nu}(z)\equiv z^{-\nu} J_{\nu} (z)\;,\; U(0)=(\frac{1}{2})^{\nu}
  \frac{1}{\Gamma(\nu+1)},
\end{equation}
and thus $U_{\nu}$ is invertible at least in the small vicinity
of origin. On the other hand, $U_{\nu}$ definitely has zeros which
coincide with zeros $z_{(\nu)k}$ of Bessel function
$J_{\nu}(z_{(\nu)k})=0$, except $z=0$ point. According to
(\ref{dress}-\ref{dress2}), this brings the following
interpretation of $m^2 \neq 0$ deformations of traceless higher
spin models. Consider such fields that operator
$\mu\rho(q\mapsto\frac{\pa}{\pa p}, d)$ (corresponding to the
argument of Bessel functions) is not equal to any of Bessel
functions zeros $z_{(s+\gamma)k}$. Then the $m^2 \neq 0$ theory
is identified with $m^2=0$ one by means of invertible nonlocal
change of variables of the form \re{dress1} with dressing
$U(\rho)$ given by (\ref{dress4},\ref{dress5}).

On the other hand, the dressing becomes non-invertible if the
dressing operator is zero on $h$, $\,U h(x,p)=0$, and it is clear
that zeros of $U$ are simultaneously the solutions of full
equations of motions. Therefore, the map from $m^2\neq 0$ to
$m^2=0$ models may be non-bijective, it degenerates e.g. on the
subspace of solutions for
\begin{equation}\label{dressbr}
D h(x,p)\equiv   \mu\rho(q\mapsto\frac{\pa}{\pa p},
d)h(x,p)=z_{(s+\gamma)k} h(x,p)
\;,\;J_{s+\gamma}(z_{(s+\gamma)k})=0,
\end{equation}
corresponding to zeros of Bessels functions. To illustrate that
this class of solutions should not be neglected let us look for
them in the form of "plane waves"
\begin{equation}\label{pwave}
  h_{r,b}(x,p)=exp(ix^a r_a +p_a b^a),\;\;r_a=const\,,\,b^a=const,
\end{equation}
the operator $D$ \re{dressbr} acts on this function by
multiplying it by $(-\frac{m^2}{r^2})^{\frac{1}{2}} \rho(b,ir)$
and therefore \re{dressbr} is satisfied provided
\begin{equation}\label{pwave2}
(\frac{m^2}{-r^2})^{\frac{1}{2}} \rho(b,ir)=[ -\frac{m^2}{r^2}
((rb)^2-r^2 b^2)]^{\frac{1}{2}} =z_{(s+\gamma)k},
\end{equation}
"Massive" momenta $r$ correspond to $r^2=-M^2$. For massive
momenta, one may consider $r$-transversal space-like
"polarizations" $b$, then the formula \re{pwave2} determines the
"spin spectrum"
\begin{equation}\label{pwave4}
\begin{array}{c}
 (r\,b)=0\,,\, b^2=m^{-2}S_k^2\,,\,S_k \in {\bf R}
  \\ \\
 S_k=z_{(s+\gamma)k},
\end{array}
\end{equation}
governed by real zeros of Bessel function $J_{s+\gamma}$.

We see $m\neq 0$ models may possess new phenomena as compared to
their $m=0$ cousins, therefore it may be worth keeping these
theories in their original basis, where they are nonlocal.
Anyway, $m \neq 0$ models are as unique as their $m=0$ limits and
are worth studying. As argued in the introduction,
\textit{corresponding gauge fields may transfer first-order
interactions of massive point particles}. We will call the
deformed model, corresponding to $\{n,\varsigma\}$
\textit{"deformed traceless spin-($2n+\varsigma$) theory"}.

Now analyze $d=2$ case. The results are quite the same, but the
analysis is slightly different. The point is that now
$\gamma=-\frac{1}{2}<0$ and thus (see
(\ref{baseq225},\ref{bess2})) solutions to the Bessel's equation
with negative index are allowed. But the only new solution which
is not covered by the formula \re{baseq10} contains a single point
along $L_{iterate}$,$\;l=0, \nu=-\frac{1}{2}$ and may be described
by general formula \re{baseq10} as well. This is the solution
describing the lowest spin, $s=0$, and thereby it reduces to a
constant (i.e., arbitrary function of $\Box$) in the $m=0$ limit.
As in general case, all other solutions which contain finite
number of terms in $l$, are obtained from our main family
\re{baseq10} by $\nu \mapsto -\nu $ change, but they are not
analytic in $\rho^2,\rho'^2$.

Let us make a technical remark. Due to invariance w.r.t.
$a$-transformations, the actions \re{Saction},\re{baseq10} depend
only upon the special traceless combinations of $h^{m(k)}$,
described in the Appendix (one traceless tensor for each rank
$s$). Specifically, $h(x,p)$ may be represented as
\begin{equation}\label{hexp}
h(x,p)=\varphi(x,p)+(p^2+m^2)\chi(x,p),
\end{equation}
where $\chi(x,p)$ is arbitrary power series in $p_m$ while
$\varphi$ is a traceless power series:
\begin{equation}\label{hexp2}
\varphi(x,p)=\sum \limits_{s=0}^{\infty} \varphi^{m(s)}(x)
p_{m_1}...p_{m_s} \;;\;{\varphi_n}^{nm(s-2)}=0.
\end{equation}
Then it is clear the action may be written in terms of $\varphi$
by making substitution $h(x,p)\mapsto \varphi(x,p)$ in the action
\re{Saction}, after the substitution, the terms $q^2,q'^2$ in the
generating function $P(q,q')$ \re{baseq10} may be dropped as they
give vanishing contributions. This simplify the form of
$\sigma\mapsto -(qd)^2, \sigma' \mapsto -(q'd)^2$. In this basis,
gauge transformations of $\varphi(x)^{m(s)}$ appear to depend
only upon the special traceless parts of $\epsilon$
(\ref{gtra1},\ref{etr1},\ref{gtra}) and entangle
$\varphi$-components of all ranks, with $m^2$ playing the role of
entanglement magnitude. Formally, they may be disentangled by
"undressing" the $m^2\neq 0$ fields according to
(\ref{dress1},\ref{dress4},\ref{dress5}), but this transformation
is highly nonlocal and is not well-defined always, e.g., it
degenerates on the massive fields with "spin spectrum" governed by
zeros of Bessel functions (\ref{pwave}-\ref{pwave4}).

\section{Conformal invariance of gauge transformations at $m=0$.} \label{s3}

Here we demonstrate the invariance, in the $m=0$ case, of the
gauge transformations \re{lintotal} w.r.t infinite-dimensional
algebra which contains the conformal algebra of $d$-dimensional
Minkowski space $so(d,2)$ as its maximal finite-dimensional
subalgebra. The conformal algebra transforms every fixed rank
tensor into itself, while general infinite-dimensional
transformation mix all ranks.

The proof employs essentially the origin of gauge fields as
background fields of the particle's theory, as their gauge
transformations are formulated directly in terms of the particle
phase space \re{total}.

The proof will be simple, but to exhibit the simplicity we have
to start from rather general facts. Suppose we have a Lie algebra
$g$ of a group $G$ acting on the manifold $M$:
$Z=Z+g(Z)+O(g)\;;\;Z \in M$. Let $Z_v \in M$ be some point
("vacuum") and $g_v \subset g$ its stability subalgebra,
\begin{equation}g_v (Z_v) =0. \end{equation} It is well known
that there is representation $T_{g_v}$ of the stability
subalgebra in the tangent space to the point $Z_v$, $T(Z_v)$:
\begin{equation}
T_{g_v} Y=(\frac{\pa g_v}{\pa Z}|_{Z=Z_v}) Y; Y \in T(Z_v).
\end{equation}
Consider the shift of the vacuum $Z_v$ w.r.t. general element of
the algebra $g$, as a function of $g$ with values in $T(Z_v)$:
$\delta_g\,Z_v=g(Z_v)\equiv R_v(g)$. Then as it is clear that
($[,]$ is Lie algebra commutator)
\begin{equation}
 T_{g_v} g(Z_v) =[g_v,g](Z_v),
\end{equation}
$R_v(g)\equiv g(Z_v)$ satisfies the equation
\begin{equation} \label{ttt} T_{g_v} R_v(g)=R_v([g_v,g]).\end{equation}
This means that general variation of the vacuum $R_v(g)$ is
covariant w.r.t. stability subalgebra transformations in the
tangent space.

Now specify the "manifold" $M$ and the algebra of transformations
$g$. The "manifold" is the space of all Hamiltonians $H(x,p)$
(understood as power series in momenta). Recall the full gauge
transformations in the space of all Hamiltonians \re{total}:
\begin{equation}\label{total1}
\delta_{(\epsilon,a)} H(x,p)= a(x,p) H(x,p)+\{ \epsilon,H(x,p)\}.
\end{equation}
This transformations are easily seen to form an
infinite-dimensional algebra $g$, isomorphic to the semidirect
product of all canonical transformations $\epsilon$ to an abelian
ideal of "hyperWeyl" transformations $a$: \be  \label{ctrs} \ba
[\delta_{(\e_1,a_1)} ,
\delta_{(\e_2,a_2)} ] H =  \delta_{(\e_3,a_3)}  H\\ \\
\e_3= \{\e_1,\e_2\} \;\;,\;\; a_3=\{\e_1,a_2\} - \{\e_2,a_1\}. \ea
\ee The stability subalgebra $g_v$ of a point $H_v$ consists of
all parameters $\epsilon_v,a_v$, satisfying the equation
\begin{equation}\label{total2}
 a_v(x,p) H_v+\{ \epsilon_v,H_v(x,p)\}=0.
\end{equation}
It is worth pointing out that the stability subalgebra has direct
physical interpretation of \textit{global symmetries} algebra of
the point particle with the Hamiltonian $H_v$. Indeed, it is easy
to see every canonical transformation $\epsilon_v$ maps the
equations of motion of the particle with Hamiltonian $H_v$ into
themselves (remember that particle's dynamics is bound to the
constraint surface $H=0$).

Now specify the function $R_v(g)$. It is given by general
variation of the vacuum Hamiltonian,
\begin{equation}\label{ttt1}
  R_v(g)\equiv R_v(\epsilon,a)= a(x,p) H_v+\{ \epsilon(x,p),H_v\}
\end{equation}
Of our main concern is the covariance property \re{ttt}. As the
Lie algebra action in the space of all $H$ is linear, the $T_v$
representation coincides with $g$ action (provided tangent space
is canonically identified with original linear space of $H$), so
we obtain
\begin{equation}\label{ttt2}
 \begin{array}{c}
  (a_v+\{\epsilon_v,.\})R_v(\epsilon,a)\equiv(a_v+\{\epsilon_v,.\})(
a H_v+\{ \epsilon,H_v\}) \\ \\
  =R_v(\{\epsilon_v,\epsilon\},\{\epsilon_v,a
  \}-\{\epsilon,a_v\}),
 \end{array}
\end{equation}
which of course may be checked by direct calculation.

The next step is the appreciation of the fact that general gauge
variation of the vacuum $R_v(g)$ is nothing but the linearized
gauge transformation for the fluctuation $h$ of general
Hamiltonian $H=H_v +h$ around the vacuum $H_v$,
\begin{equation}\label{ttt3}
  \delta_{(\epsilon,a)} h(x,p) =\delta_{(\epsilon,a)} H_v =R_v({\epsilon,a}).
\end{equation}
Thus the linearized gauge transformations \re{ttt3} possess
covariance w.r.t. global symmetry group $g_v$:
\begin{equation}\label{ttt4}
  (a_v+\{\epsilon_v,.\})\delta_{(\epsilon,a)} h(x,p)=\delta_{(\{\epsilon_v,\epsilon\},\{\epsilon_v,a
  \}-\{\epsilon,a_v\})} h(x,p).
\end{equation}
This means that the infinitesimal global symmetry transformations
\begin{equation}\label{ttt5}
\delta_{(\epsilon_v, a_v)} h =(a_v+\{\epsilon_v,.\}) h
\end{equation}
result in new $h$ that obeys \textit{exactly the same} gauge laws
but with transformed gauge parameters
\begin{equation}\label{ttt6}
\delta_{\epsilon_v, a_v} \epsilon
=\{\epsilon_v,\epsilon\}\;;\;\delta_{(\epsilon_v, a_v)}
a=\{\epsilon_v,a  \}-\{\epsilon,a_v\}.
\end{equation}
As we are always able to redefine the gauge parameters, all this
is equivalent to the statement that \textit{while $h$ changes, the
gauge transformations do not change}.

One more important property concerns the action on $h$ of
\textit{trivial global symmetries} $(\epsilon_v,a_v \in
g_{triv})$ of the form
\begin{equation}\label{trivsym}
  \epsilon^{(triv)}_v = \mu(x,p) \, H_v\;,\; a_v^{(triv)}=- \{\mu(x,p),H_v\}
\end{equation}
with $\epsilon_v$ vanishing on the constraint surface, where $\mu$
is an arbitrary power series in $p$. One may check that
\begin{equation}\label{trivsym1}
  \delta_{(\epsilon^{(triv)}_v , a_v^{(triv}))} h =R_v(-\mu h,
  \{\mu,h\})
\end{equation}
and thereby trivial global symmetries act on $h$ as some
$h$-dependent gauge transformations. It is easy to check that
$g_v^{triv}$ form an ideal in $g_v$. So, the space of gauge
invariants of $h$ (and, needless to say, the physical phase space
of the particle) acquires action of the \textit{algebra of
obsevables} $g_o$ which is defined as a factor-algebra
\begin{equation}\label{ao}
g_o \equiv g_v/g_{triv}.
\end{equation}

 Now let us apply all these matters to the Hamiltonian
of the massless particle on Minkowski space
\begin{equation}
H_v=\frac{1}{2}p^2
\end{equation}
It is well-known the massless
particle's theory possesses conformal invariance, and our
derivation allows one to transfer this invariance to the gauge
transformations of traceless higher spin theories.

 Consider the canonical generators of conformal
transformations of $d$-dimensional Minkowski metric on the
particle's phase space:
\begin{equation}\label{conft}
 \begin{array}{c}
  \epsilon_{c}=k^{ab} M_{ab} + b^a P_a + f^a K_a +f D; \\
M_{ab}=x_a p_b -x_b p_a\;;\; P_a=p_a\;;\;
K_a=x^2p_a-2(xp)x_a\;;\; D=(xp).\end{array}
\end{equation}
Here $k^{ab},b^a,f^a,f$ are parameters for infinitesimal Lorentz
transformations, translations, special conformal transformations
and dilations, respectively. By their very definition, all these
generators either leave invariant the Hamiltonian of the massless
particle $H_v=\frac{1}{2}p^2$ (Poincare generators $M_{ab},P_a$)
or scale it by a function of $x$ (special conformal $K_a$ and
dilations $D$):
\begin{equation}\label{conft1}
\{\epsilon_{c},p^2\}=(2D-4f^a x_a) p^2 \equiv -a_{c}p^2.
\end{equation}
Note that any product of $\epsilon_{c}$ possesses this property
either:
\begin{equation}\label{conft11}
\begin{array}{c}
 \Upsilon_t=\epsilon_{(1)c} \;\epsilon_{(2)c}...\epsilon_{(t)c}\Rightarrow
 \\ \\
\{ \Upsilon_t,p^2 \}=-(a_{(1)c}\;
\epsilon_{(2)c}...\epsilon_{(t)c}
+\;...\;+\epsilon_{(1)c}...\epsilon_{(t-1)c}\; a_{(t)c})\equiv
-A_t p^2.
\end{array}
\end{equation}
Comparing the last equation with \re{total2} we see that pairs
$\Upsilon_t, A_t$ are the elements of the stability subalgebra of
the vacuum, or the global symmetries (of the particle). Moreover,
the linear space of all $\Upsilon_t,A_t, t=1,2,3,...$ is the
infinite-dimensional Lie algebra w.r.t. to the composition law
\re{ctrs}, while $\Upsilon_1,A_1$ is the original conformal
algebra. Thus, at least $\Upsilon_t,A_t, t=1,2,3,...$ form a
subalgebra $\bar{g}_v$ of the total symmetry algebra $g_v$, but
actually one can show that any global symmetry (which has a form
of power series in momenta) is represented as a combination of
$\Upsilon_t,A_t$, so $\bar{g}_v =g_v$. Among global symmetries,
there are \textit{trivial} ones $(\epsilon_v,a_v \in g_{triv})$ of
the form \re{trivsym}
\begin{equation}\label{trivsym2}
  \epsilon_v =\frac{1}{2}p^2 \mu(x,p)\;,\; a_v=-p\pa_x \mu(x,p),
\end{equation}
The algebra of observables defined as \re{ao} is an infinite
dimensional algebra isomorphic to a contraction of some conformal
higher spin algebra of the type proposed by Fradkin and Linetsky
in $d=3,4$ \cite{Fradkin:1989xt,Fradkin:1989yd}. It may be shown
that, if one considers the quantum massless particle, its algebra
of observables deforms to non-contracted conformal higher spin
algebra \cite{preparation}.

According to the above reasoning, all transformations from $g_o$
\re{ttt5} present a symmetry of the gauge transformations
\re{lintotal} in the case $m^2=0$. The finite-dimensional
conformal subalgebra preserves the subspace of every $s$-th degree
in momenta and thereby acts separately on each spin-$s$ model,
while higher transformations mix all spins.

Now if the gauge transformations were determined the wave
equation for $h$ uniquely, then one could state that while $h$
changes, the wave equation does not change, so then
transformations \re{ttt5} would transform the space of solutions
of the wave equation into itself, i.e. present a symmetry of the
free wave equation for $h$. However, our solution for wave
equation were unique only under requirements of Poincar\'e
invariance, while special conformal transformations $K_a$ may
break it, so for analyzing the conformal invariance of our wave
equations more information is required.

\section{Conclusion.}

Starting from gauge transformations, induced by the first-order
point particle-symmetric tensors interactions, we have constructed
Poincar\'e- and gauge-invariant free actions for traceless
symmetric tensor fields (which should not be confused with unitary
\textit{double-traceless} higher spin theories \cite{fronsdal1}).
These actions set some dynamics for these fields which thereby
may mediate point particles interactions. The typical processes
may look just like those in classical electrodynamics: there
exists a free field "F" which propagates through space-time
according to its equations of motion, and there are sources
localized on point particle world lines. To study interaction of
two particles, "A" and "B",  one has to solve the equations for
free gauge fields "F" with a source formed by "A" and then to
study the motion of "B" in the field "F", and vice versa. This is
one of interesting tasks to be studied in future.

One of the manifestations of importance of these models is their
\textit{uniqueness}. For each spin $s$, there is just one family
of theories parameterized by the particle's mass $m$. At the point
$m=0$, the theories reduce to local higher derivative
scale-covariant models with actions of $2s$-th order in
derivatives. In $d=4$, these models have been described by Fradkin
and Tseytlin ("pure spin" models) \cite{FT} and their
supersymmetric cubic interactions were elaborated by Fradkin and
Linetsky \cite{FL1,FL2}. It may be interesting to note that the
gauge transformations (3.5) in \cite{FL1} look identically to our
\re{lintotal1} (at $m=0$), but there they are not linked to any
first-order interaction.

The "pure spin" models were claimed to be conformally
invariant\cite{FT,FL1,FL2}. Yet, we have not found a simple way to
analyze the  covariance properties of traceless higher spin
theories w.r.t conformal group. Perhaps, the most appropriate way
to study conformal invariance in the models is to reformulate
them in $2T$-physics framework advocated by Bars
\cite{Bars:2001um}. Besides, we have presented the action of an
infinite dimensional Lie algebra, (which contains full conformal
algebra $so(d,2)$ of $d$-dimensional Minkowski space as a maximal
finite-dimensional subalgebra) on the gauge fields, this action
appears to leave gauge transformation intact. This algebra may be
deformed to the "conformal higher spin algebra in $d$-dimensions"
analogous to that studied in \cite{Fradkin:1989xt,Fradkin:1989yd},
which is the same as "higher spin algebra" in $d+1$ dimensions
\cite{preparation,Konstein:2000bi}, that may present interest in
view of $AdS/CFT$ correspondence. The origin of gauge fields as
background fields in the particle's theory is basic in our
derivation.

As $m^2\neq 0$, the theories become essentially nonlocal as it is
manifested by inverse powers of $\Box$ up to an infinite order in
the generating function \re{baseq10}, and entangle traceless
tensors of all spins. Although these peculiar properties may be
cured by a nonlocal "undressing" change of variables
(\ref{dress1},\ref{dress4},\ref{dress5}) which maps $m \neq 0$
models to their $m=0$ counterparts of the same spin, it may be
worth dealing with $m\neq 0$ case in the original basis (where
they are nonlocal) as otherwise one may throw away some
potentially important solutions. We have presented a class of
solutions of this type, with "spin spectrum" governed by zeros of
Bessel functions (\ref{pwave}-\ref{pwave4}). Anyway, "deformed
traceless higher spin theories" are interesting to study as they
are, in a sense, unique and able to mediate interactions of
massive point particles.

Besides studying new interactions of point particles, another
immediate task is a nonlinear theory of traceless fields. Indeed,
as the quadratic actions do exist and the nonlinear gauge
transformations are known \re{total}, it is worth examining the
perturbative solution for the nonlinear action
\cite{preparation1}. If the solution exists it may present a new
gauge theory generalizing conformal gravity and describing
interactions of infinite number of traceless tensor fields.

\section*{Acknowledgement}
The author appreciates discussions on diverse aspects of higher
spin theories with R. Metsaev, A. Sharapov, A. Tseytlin, I. Tyutin
and M. Vasiliev. The author is grateful to Professor Bernard de
Wit for invitation and hospitality at  Spinoza Institute,
Utrecht. The work is supported by INTAS grant No. YSF-00-149, RFBR
grant 99-02-17916 and scientific schools RFBR grant 00-15-96566.

\vspace{5mm}
\appendix{\bf  Appendix. Gauge transformations for traceless tensors.}
\vspace{5mm}

\noindent Here we rewrite gauge transformations \re{lintotal}
\begin{equation}\label{lintotalap}
\delta h(x,p)= \frac{1}{2} a(x,p)(p^2+m^2) + p_m \eta^{mn}
\partial_n \epsilon (x,p)
\end{equation}
in terms of $a$-invariant traceless tensors.

We need some simple tools to handle traces of tensor coefficients
of arbitrary functions. Let us note that, given any function
$$F(x,p)=\sum\limits_{k=0}^\infty F^{m(k)} (x)
p_{m_1}...p_{m_k},$$ one can unambiguously represent it in the
form \be \label{decom} F(x,p)= \sum\limits_{l=0}^\infty
\sum\limits_{k=0}^\infty F_{(l)}^{m(k)} (p^2)^l p_{m_1}...p_{m_k},
\ee where $F_{(l)}^{m(k)}$ are traceless,
${{F_{(l)}}{}_n}^{nm(k-2)}=0$. This is easily done by decomposing
each $F^{m(k)}$ to its traceless part and the traces
$F^{m(k)}=F_{(0)}^{m(k)}+\eta^{m(2)}F_{(1)}^{m(k-2)}
+\eta^{m(2)}\, \eta^{m(2)} F_{(2)}^{m(k-4)} +...$, then summing up
the power series by momenta and noting that the trace parts give
the powers of $p^2$. The decomposition \re{decom} is then
rewritten as \be F(x,p)= \sum\limits_{k=0}^\infty F^{m(k)} (p^2)
p_{m_1}...p_{m_k}, \ee where $F^{m(k)} (p^2)=
\sum\limits_{l=0}^\infty F_{(l)}^{m(k)} (p^2)^l. $ Decomposing
the power series $F^{m(k)} (\sigma)$ at the point $\sigma=-m^2$
one gets \be \label{decom2}\begin{array}{c}
 F(x,p)=\sum\limits_{k=0}^\infty F_{[-m^2]}^{m(k)} (p^2+m^2)
p_{m_1}...p_{m_k}= \\ \\\sum\limits_{l=0}^\infty
\sum\limits_{k=0}^\infty F_{[-m^2](l)}^{m(k)} (p^2+m^2)^l
p_{m_1}...p_{m_k}= \sum\limits_{l=0}^\infty
F_{[-m^2](l)}(p^2+m^2)^l,
\end{array}
\ee where the power series $F_{[-m^2](l)}$ contain only traceless
coefficients. Given $m^2$, we will say that the $F_{[-m^2](0)}$
term is the traceless part of the function $F(x,p)$ and the
first, second and further traces of $F$ are represented by
$F_{[-m^2](l)} (p^2+m^2)^l$ with $l=1,2,...$ forming altogether
the traceful part of $F$. The function is traceless if it is
equal to its traceless part. In this sense each coefficient
$F_{[-m^2](l)}$ is a traceless function.

Now represent all the entries of the gauge transformation laws
\re{lintotalap} in the form \re{decom2} to get \be \label{Glaws2}
\de \sum\limits_{l=0}^{\infty}  h_{[-m^2]\,(l)}\, (p^2+m^2)^l =
\sum\limits_{l=0}^{\infty} \left\{\frac{1}{2}a_{[-m^2]\,(l)}
(p^2+m^2)^{l+1} + p^m \pa_m \e_{[-m^2]\,(l)} (p^2+m^2)^l\right\}
\ee wherefrom it is seen that all the traces of $h$ may be gauged
away by $a$-transformations. In fact, the very destination of $a$
is to gauge away the traces of $h$. It is worth noting that the
traceful part of $\e$ is already contained in $a$ as the gauge
transformations \re{Glaws2} do not change if one redefines $\e,a$
according to \be \de \e= \frac{1}{2}(p^2+m^2)\nu\,,\,\de a= - p^m
\pa_m \nu. \ee Therefore without loosing a generality one may set
$\e$ traceless \be \label{etr1}\e= \e_{[-m^2]0} \equiv \ve
=\sum\limits_{k=0}^{\infty} \ve^{m(k)} p_{m_1}...p_{m_k}\,;\,
{\ve_n}^{nm(k-2)}=0. \ee

For any action ${\cal A}[h]$ invariant w.r.t. gauge
transformations \re{lintotalap}, $h$ should enter ${\cal A}[h]$ in
$a$- and $c$-invariant combinations only as far as $a$
transformations are purely algebraic. It is easy to see that the
only $a$-invariant is the traceless function
 \be \label{5} \varphi=h_{[-m^2](0)}. \ee

It is easy to derive the $\ve$ transformation laws for the
coefficients of $\varphi$ which read \be \label{gtra}  \de
\varphi^{\,m(s)} =( {\mbox{Traceless part of}}\; \pa^{m}
\ve^{m(s-1)}) -m^2\, \frac{s+1}{2s+d}\, \pa_n {\ve^{nm(s)}}.\ee
For $m^2=0$, these are the gauge transformations of conformal
higher spin theories, which are seen to decay in independent
subsystems described in terms of rank-$s$ traceless tensor and
rank-$(s-1)$ traceless parameter. For $m^2 \neq 0$, as it is
proved in the main text, for each spin $s$ there exists the
deformed traceless higher spin theory, which  reduces to
traceless spin-$s$ model in the limit $m^2\rightarrow 0$. The
deformed theory is nonlocal, with nonlocality being measured by
$\frac{m^2}{\Box}$. We now see from \re{gtra} that traceless
tensors of all ranks get entangled by gauge transformations in
$m^2 \neq 0$ case. In this sense, $m^2$ exhibits itself also as
an entanglement magnitude.

\end{document}